\documentclass{llncs}

\usepackage[utf8]{inputenc}
\usepackage{amsmath,amssymb}
\usepackage{stmaryrd}
\usepackage{hyperref}
\usepackage{url}
\usepackage{fancyhdr}

\begin{document}

\renewcommand{\topfraction}{0.9}
\renewcommand{\bottomfraction}{0.9}
\setlength{\abovedisplayskip}{0pt}
\setlength{\belowdisplayskip}{0pt}
\setlength{\abovedisplayshortskip}{0pt}
\setlength{\belowdisplayshortskip}{0pt}

\setlength{\itemsep}{0pt}
\setlength{\parsep}{0pt}
\setlength{\parskip}{0pt}

\title{D2d --- XML for Authors}
\subtitle{(Technical Report -- Bad Honnef 2013)}

\author{Markus Lepper \inst1
\and
Baltasar Tranc{\'o}n y Widemann \inst1
}

\institute{
{\tt<}semantics{\tt/>} GmbH, Berlin
\\ \texttt{post@markuslepper.eu}
}

\maketitle

% \emptysequence
\newcommand{\xes}{\ensuremath{\langle\rangle}}

\newcommand{\xtail}{\ensuremath{\blacktriangleleft}}
\newcommand{\xhead}{\ensuremath{\blacktriangleright}}
\newcommand{\chop}{\ensuremath{\smallfrown}}
\newcommand{\hateq}{\ensuremath{\stackrel{\wedge}{=}}}

\newcommand{\fun}{\ensuremath{\rightarrow}}
\newcommand{\inj}{\ensuremath{\rightarrow\kern-1.2em\rightarrow}}
\newcommand{\pinj}{\ensuremath{\rightarrow\kern-1.2em\nrightarrow}}
\newcommand{\pfun}{\ensuremath{\nrightarrow}}
\newcommand{\pffun}{\ensuremath{\rightarrow\kern-2.75ex\arrownot\,\arrownot\kern2.25ex}}
\newcommand{\rel}{\ensuremath{\leftrightarrow}}
\newcommand{\finrel}{\ensuremath{\leftrightarrow\kern-2.5ex\arrownot\,\arrownot\kern2ex}}
\newcommand{\xpower}{\ensuremath{\mathop{\mathbb{P}{}}}}
\newcommand{\xfinpower}{\ensuremath{\mathop{\mathbb{F}{}}}}
\newcommand{\dom}{\ensuremath{\mathop{{}\mathsf{dom}}}}
\newcommand{\ran}{\ensuremath{\mathop{{}\mathsf{ran}}}}
\newcommand{\nat}{\ensuremath{\mathbb{N}}}
\newcommand{\compose}{\ensuremath{\fatsemi}}
\newcommand{\limg}{\ensuremath{\mathopen{(\kern-0.4ex|}}}
\newcommand{\rimg}{\ensuremath{\mathclose{|\kern-0.4ex)}}}
\newcommand{\rinv}{\ensuremath{^\sim}}
\newcommand{\domres}{\ensuremath{\triangleleft}}
\newcommand{\ndomres}{\ensuremath{\triangleleft\kern-2.1ex-}}

\newcommand{\xlsem}{\ensuremath{\big[\kern-0.6em\big[~}} 
\newcommand{\xrsem}{\ensuremath{~\big]\kern-0.6em\big]}} 

\newcommand{\xDef}{\ensuremath{\stackrel{\mathit{def}}{\Longleftrightarrow}}}

\newcommand{\xrule}[1]{\mbox{~\small\bf(#1)}}

\newcommand{\makeobjectTT}[1]%
{\expandafter\newcommand\csname x#1\endcsname{\ensuremath{\mathtt{#1}}}}
\newcommand{\makeobjectIT}[1]%
{\expandafter\newcommand\csname x#1\endcsname{\ensuremath{\mathit{#1}}}}
\newcommand{\makeobjectSF}[1]%
{\expandafter\newcommand\csname x#1\endcsname{\ensuremath{\mathsf{#1}}}}

\newcommand{\xfrac}[2]{\frac{\begin{array}{c}#1\end{array}}%
{\begin{array}{c}#2\end{array}}}

% ----------------------------------------------------------------------

\newcommand{\ddd}{\textsf{d2d}}
\newcommand{\Ddd}{\textsf{D2d}}

\newcommand{\xop}[1]{\mbox{\,\texttt{\large #1}\,}}  %% mathop infixop ?? FIXME

\makeobjectSF{defs}

\makeobjectIT{Definition}
\makeobjectTT{chars}
\makeobjectTT{tags}
\makeobjectTT{comment}
\makeobjectTT{OPEN}
\makeobjectTT{CLOSE}
\makeobjectTT{EMPTY}
\makeobjectTT{eof}
\makeobjectSF{tokenize}
\makeobjectTT{node}
\makeobjectTT{perm}
\makeobjectTT{missing}
\makeobjectTT{skipped}
\makeobjectTT{skip}
\newcommand{\Xchars}{\ensuremath{\mathtt{\#chars}}}
\newcommand{\Xeof}{\ensuremath{\mathtt{\#eof}}}
\makeobjectSF{ascend}
\makeobjectSF{descend}
\newcommand{\texttree}{\ensuremath{\mathsf{text2tree}}}
\makeobjectSF{translate}
\makeobjectSF{first}
\makeobjectSF{potEps}
\makeobjectSF{boolean}
\makeobjectSF{parseM}
\makeobjectSF{parse}

\newcommand{\XSL}{\mbox{\texttt{/}}}
% stmaryd:
\newcommand{\XSLC}{\ensuremath{\fatslash}}
\newcommand{\XNL}{\ensuremath{{}^N_L}}
\newcommand{\XH}{\ensuremath{\mathtt{\#}}}
\newcommand{\XB}{\mbox{\textvisiblespace}}
\newcommand{\xtag}{\ensuremath{\mathtt{someTag}}}
\newcommand{\xcp}[2]{\ensuremath{\mbox{'\texttt{#1}'}\mapsto\mbox{'\texttt{#2}'}}}

% --------------------------------------------------------------------------------
\section{Introduction, Principles of \ddd}

XML is a de facto standard for the encoding of semi-structured text corpora.
Its practical application ranges from mere technical configuration data
to web sites with entertaining contents.

Both notions, ``XML'' and ``text'', stand here
for very different things: On one side the organization of the internal
computable text model as a tree structure with its standardized
update and retrieval methods (``W3C DOM''), and the family of tools operating on these
(implementing ``XSLT'', ``XQuery'', etc.).
On the other side its external representations: unicode text files containing a lot
of ``angled braces'', their decoding ruled by a historically grown 
collection of hardly understandable, non-compositional quoting rules.

All this hinders the creation of XML encoded texts 
in the \emph{creative flow of authoring}. Neither are syntax controlled editors,
which support tagging by menu driven selections,
auto completion, automated coloring and indenting,
a solution for all those authors which are used to ``writing''
as a creative, flowing, intuitive and intimate process, in \emph{direct}
contact with that mysterious thing called text.

So far XML appears inappropriate for this kind of authoring situation.
Nevertheless often its use is highly desirable:
Technical documentations, cookbooks, screen plays, song lyrics, 
scientific analyses, even multi-volume fantasy novels
can profit extraordinarily by only little interspersed mark-up.

This is the starting point of the ``\ddd'' project. It stands
for ``directly to document'' or ``direct document denotation'',
hence pronounced ``triple dee'',  and it tries to close this gap.
It is both, a text format which realizes XML mark-up in a very unobtrusive 
way, and it is a software system which implements parsing, translating,
parser definitions, documentation, user guidance, etc.

It is based on a simple idea, the realization of which turned out to be
surprisingly complex, and has been driven on by the authors for now
more than ten years.\footnote{A rather early version is described in
\cite{ltw01b}; full documentation can be found at \cite{metatoolsdoc}.}
The main characteristics are:
\begin{itemize}
\item
Simple way of writing and good readability by humans (without the need for 
any dedicated tool) as well as by machines.
\item
All tags marked with one single, user-defined character.
\item 
Inference of (nearly all) closing tags.
\item
Inference also of opening tags by a second, character-based level of parsers,
used for small, highly structured data entities, interspersed in flow text.
\item
Support of standard text format definition formats (e.g.\ W3C DTD).
\item
Own language for text format definition (required at least for the
character level parser definition). It employs free rewriting for 
parametrization and re-use of modules, and multi-lingual user documentation.
\end{itemize}

So \ddd{} is a concept, a format and a software system which addresses domain experts
and enables them to write XML compatible texts and potentially opens to them the
whole world of XML based processing.
\Ddd{} has been sucessfully employed in the very diverse fields of technical
documentation, book keeping, web content creation, interactive music theory, etc.

% --------------------------------------------------------------------------------
\section{The \ddd{} Parsing Process}

% --------------------------------------------------------------------------------
\subsection{Principles of the \ddd{} Parsing Process}

The process of reading a text file, 
interpreting it as conforming to particular text format definition in the \ddd{} format,
and constructing the corresponding internal model, is called ``\Ddd{} parsing''.
This model can be written out as an external representation according to
the XML standard \cite{xml11}.
The implementation of the \ddd{} tool also allows to process this model directly,
e.g.\ using a collection of XSLT rules to derive other models to be written out.
The parsing process is controlled by the chosen root element.
Element definitions can have tagged or character content models:

In the \textbf{tagged} case, the resulting sub-tree of the model is constructed
according to the tags appearing in the input text. Tags are marked by
a single user-defined lead-in character, which defaults to ``\texttt{\#}''.
Closing tags can in most cases be omitted, because the parser uses
simply LL(1) strategy: Whenever an opening tag can be accepted on some
currently open higher level in the result tree, all intervening closing
tags are supplied automatically.
Nevertheless explicit closing tags may be added to resolve ambiguities, to close more
than one stack level, or to increase the readability of the source.

\textbf{Character} based parsers accept plain character data. All tagging is 
added automatically as defined in the applicable parser rules.
The basic strategy is a non-deterministic longest match. 
Thus this mechanism performs well
for short input data, e.g.\ some ten lines of MathML.
In practice this covers most instances of structured entities,
interspersed in flow text. Besides, non-deterministic rules are much easier
definable by computer language laymen.

% --------------------------------------------------------------------------------
\subsection{File Sections}

% -----------------------------------------------------------------------
\begin{table}[tp]
  \begin{align*}
  \begin{gathered}
C = \mbox{\textit{//~~allowed input characters }}\\
I = \mbox{\textit{//~~allowed identifiers }}\\
\begin{array}{lcl}
    D &::=& \xchars(C^\star)~~|~~\xcomment(C^\star)~~|~~\xeof \\
&|& \mathtt{error}(C^\star)~~|~~\mathtt{warning}(C^\star) \\
&|& \xOPEN_{i:I}~~|~~\xCLOSE_{i:I}~~|~~\xCLOSE^F_{i:(I\cup\{\_\})} 
~~|~~\xEMPTY_{i:I}~~|~~\xEMPTY^F_{i:I}\\
\end{array}\\
  \xtokenize : C^\star \fun D^\star \\
\_\leadsto\_  : (D^\star \times C^\star \times (C\times I)^\star) \fun 
          (D^\star \times C^\star \times (C\times I)^\star) \\
  \frac{ (\xes, \alpha, \xes) \leadsto  (\delta, \xes, \xes)}
       { \xtokenize (\alpha) = \delta} \\
\mathsf{Parenths}~~:~~C~\pinj~C \\
\begin{array}{ll}
\mathsf{Parenths}~~=~~\{ & \xcp(), \xcp<>, \xcp[], \xcp\{\}, \xcp..  \xcp!!,  \\
               & \xcp\textbackslash\textbackslash, \xcp::, \xcp\$\$, 
\xcp{$\uparrow$}{$\uparrow$}
  \}
\end{array}\\[1ex]
\frac{ 
\begin{array}{c}
\alpha, \alpha'' \in C^\star~~~~\alpha''\not=\xes~~~\delta \in D^\star\\
\gamma \in (C \setminus \{\XNL\})^\star 
~~~~\gamma' \not= \_\chop \mathtt{*}\XSLC\chop \_  \\
\kappa \in \dom \mathsf{Parenths}~~\land~~\kappa' = \mathsf{Parenths}(\kappa) \\
\zeta \in (\mathtt0 | \mathtt1 | \ldots | \mathtt9),
(\mathtt0 | \mathtt1 | \ldots | \mathtt9 | \mathtt a |  \ldots | \mathtt f)^\star
~~~\land~~\zeta' = \mathtt{char(parseInt(}\zeta\mathtt{))}\\
\alpha' \in (C\setminus\{\XH, \kappa'\})^+ 
~~\land~~\alpha'\not= \_\chop \XSLC\XSLC\chop\_ 
~~\land~~\alpha'\not= \_\chop \XSLC\mathtt{*}\chop\_ 
\\
\end{array}
}
{\begin{array}{llcll}
(\delta, & \XH (\XH|\XB|\XNL)^+ \chop \alpha, \pi) & \leadsto 
                    & (\delta, & \XH\xhead\alpha, \pi) \\
(\delta, & \XSLC\XSLC\gamma\XNL  \xhead\alpha, \pi)&\leadsto 
                    & (\delta \xtail \xcomment(\gamma), & \XNL\xhead \alpha, \pi)\\
(\delta, & \XSLC\mathtt{*}\gamma'\mathtt{*}\XSLC \xhead\alpha, \pi)&\leadsto 
                    & (\delta \xtail \xcomment(\gamma'), & \alpha, \pi)\\
(\delta, & \XH\xtag(\XB|\XNL) \xhead\alpha, \pi)&\leadsto 
                    & (\delta \xtail \xOPEN_{\xtag}, & \alpha, \pi) \\
(\delta, & \XH\xtag\kappa \xhead\alpha, \pi)&\leadsto 
                  & (\delta \xtail \xOPEN_{\xtag}, & \alpha, (\kappa',\xtag)\xhead \pi) \\
(\delta, & \kappa' \xhead\alpha, (\kappa', \xtag)\xhead \pi)&\leadsto 
                    & (\delta \xtail \xCLOSE_{\xtag}, & \alpha, \pi) \\
(\delta, & \XH\xtag\XSL \xhead\alpha, \pi)&\leadsto 
                    & (\delta \xtail \xEMPTY_{\xtag}, & \alpha, \pi) \\
(\delta, & \XH\xtag\XSL\XSL\XSL \xhead\alpha, \pi)&\leadsto 
                    & (\delta \xtail \xEMPTY^F_{\xtag}, & \alpha, \pi) \\
(\delta, & \XH\xtag \xhead\alpha, \pi)&\leadsto 
                    & (\delta \xtail \xOPEN_{\xtag}, & \alpha, \pi) \\
(\delta, & \XH\zeta\XSL \xhead\alpha, \pi)&\leadsto 
                    & (\delta \xtail \xchars(\zeta'), & \alpha, \pi) \\
(\delta, & \XH\zeta \xhead\alpha, \pi)&\leadsto 
                    & (\delta \xtail \xchars(\zeta'), & \alpha, \pi) \\
(\delta, & \XH\XSL\xtag(\XB|\XNL) \xhead\alpha, \pi)&\leadsto 
                    & (\delta \xtail \xCLOSE_{\xtag}, & \alpha, \pi) \\
(\delta, & \XH\XSL\XSL\XSL\xtag(\XB|\XNL) \xhead\alpha, \pi)&\leadsto 
                    & (\delta \xtail \xCLOSE^F_{\xtag}, & \alpha, \pi) \\
(\delta, & \XH\XSL(\XB|\XNL) \xhead\alpha, \pi)&\leadsto 
                    & (\delta \xtail \xCLOSE_{\_}, & \alpha, \pi) \\
(\delta, & \XH\XSL\xtag \xhead\alpha, \pi)&\leadsto 
                    & (\delta \xtail \xCLOSE_{\xtag}, & \alpha, \pi) \\
(\delta, & \XH\XSL\XSL\XSL\xtag \xhead\alpha, \pi)&\leadsto 
                    & (\delta \xtail \xCLOSE^F_{\xtag}, & \alpha, \pi) \\
(\delta, & \XH\XSL \xhead\alpha, \pi)&\leadsto 
                    & (\delta \xtail \xCLOSE_{\_}, & \alpha, \pi) \\
(\delta, & \alpha'  \xhead\alpha, (\kappa',\_)\xhead \pi)&\leadsto 
                    &(\delta \xtail \xchars(\alpha'), &\alpha, (\kappa',\_)\xhead\pi) \\
(\delta, & \XH\xeof\xhead\alpha, \pi\not=\xes) &\leadsto
    &\multicolumn2l{(\delta \xtail \mathtt{warning}\mbox{(``pending\ parentheses'')},}\\
         &          &     & & \XH\xeof\xhead\alpha,\xes)\\
(\delta, & \XH\xeof\xhead\alpha'', \xes) &\leadsto
    &\multicolumn2l{(\delta \xtail 
                     \mathtt{warning}\mbox{(``discarding  trailing  characters'')},}\\
         &          &     & & \XH\xeof,\xes)\\
(\delta, & \XH\xeof, \xes) &\leadsto
                    &(\delta \xtail \xeof, & \xes, \xes)\\
(\delta, & \xes, \pi) &\leadsto
    &\multicolumn2l{(\delta \xtail \mathtt{error}\mbox{(``premature end of file'')},
                            \xes, \xes)}\\
\end{array}}
  \end{gathered}
  \end{align*}
\caption{Basic Data and Tokenization}
\label{tab_token}
\end{table}
% -----------------------------------------------------------------------

\newcommand{\xsp}{\XB}

Every file to be processed by the \ddd{} tool may start with 
sections containing local definitions.
These have the form 
\[ \alpha~~\XH\mathtt{d2d}~\xsp~~\mathtt{2.0}~~\xsp~~\mathtt{module}~~\mu \]
(In this paragraph  ``$\xsp$'' stands for non-empty sequences of
whitespace and newlines.)
$\alpha$ is a prefix not containing such a section, and will be discarded totally. 
This allows \ddd{} input to be contained in arbitrary documents, like e-mails etc.

$\mu$ must be a valid module definition in the ddf format (see Section~\ref{txtDdf}).
It will be parsed and the contained definitions can be used
immediately in the following text corpus.

Zero to many such local definition sections can be contained in an input file.
Finally it has to follow either
\[ \alpha~~\XH\mathtt{d2d}~\xsp~~\mathtt{2.0}~\xsp~~\mathtt{text}~~
\xsp~~\mathtt{using}~\xsp~m~\mathtt{:}~~t  \]
In this case $m$ is the name of a module, and $t$ is the name of a tag
parser definition from $m$. This is used as the topmost element
for the document structure to be parsed
and thus defines the initial state of the parsing process.
The other possibility is 
\[ \alpha~~\XH\mathtt{d2d}~\xsp~\mathtt{2.0}~\xsp~\mathtt{xslt}~
\xsp~~\mathtt{text}~~\xsp~~\mathtt{producing}~~\xsp~~m~\mathtt{:}~t  \]
In this case an XSLT source will be parsed, and the module and tag do identify
the top-level element of the output to be generated by the XSLT code.

In both cases, the rest of the file immediately after the  ``$t$'' 
is the text corpus input, fed to the \ddd{} parser, up to a final 
explicit ``\texttt{\#eof}''.

% --------------------------------------------------------------------------------
\subsection{Tokenization}

The function  $\mathsf{tokenize}$ in Table~\ref{tab_token} defines
the next step for processing the text corpus data (not the local module definitions), 
namely converting the stream of characters into a stream of tokens.
The comment lead-in character ``$\XSLC$~'' and the command character ``$\XH$''
can be re-defined by the user, and default to ``$\XSL$'' and ``\texttt{\#}'', resp.
The tokenization process is defined by
applying a \textbf{longest prefix match} discipline
to the transformation rules given for ``$\_\leadsto\_$''.
The closing and empty tags with three slashes mark those elements which are
intentionally left incomplete by the user. The reaction of the tools in these
cases is configurable.

The tokenization level supports a limited set of one character parentheses, as known
from sed's  ``\texttt{s\%...\%...\%}'' and 
\LaTeX{}'s  ``\texttt{\textbackslash verb\%...\%}'' syntax.
It is unrelated to the parser level, which can cause funny effects, but which 
nevertheless has turned out to be the cleanest way to define.

% --------------------------------------------------------------------------------
\subsection{Tag Based Parsing}

The second step, parsing, is to convert a sequences of tokens from $D^\star$ to 
a single node from $N$, as defined by the function
$\texttree()$ in Tables~\ref{tab_tagparse} and \ref{tab_tagparse_II}.\footnote{
The formulas in this tables have been published in \cite{keod2011}.}
This node represents the top-most element of the resulting document model,
which can later be shipped
out to standard XML file format, or further processed by XSLT, as soon as 
completed.

Parsing always starts in tag mode, i.e.\ looking for explicit tags
``$\xOPEN_i$'' in the token stream.
Character data is treated as if tagged with an implicit pseudo-tag
``\Xchars''. 
The tag parsing process is a stack-controlled recursive descent parser.
Whenever a tag is consumed, a new stack level from $F$ is possibly added
and the corresponding content model is made the new accepting state machine.
This is performed by the function $\xdescend()$, which delivers a new
stack prefix. Its definition is comparably simple, since 
(a) it is only called when the next tag $i$ is contained in $\xfirst(t)$, 
and (b) all \xfirst{} sets in all alternatives are disjoint.

The stack levels represent the choice points at which the parsing process can 
later be continued.
Whenever a tag (opening or closing tag) is reached which cannot be consumed in the 
current state, the stack is unwound in search for the first possibility
by functions $\xascend_O()$ and  $\xascend_C()$.
Only if such is found, all intermediate stack frames are closed, and all
material collected there is packed into \xnode{} objects.
(The third parameter of the functions is an accumulator for these;
finally their sequence is wrapped in the highest closed \xnode{} and appended
to the contents of the parent element's \xnode{}).
If some non-optional content is missing in the closed frames, 
an error message element ``\texttt{missing()}'' 
is synthesized and inserted into the
resulting model. If no such frame is found, the input is ignored,
the stack left unchanged, and a ``\texttt{skipped()}'' error message is inserted,
instead.

Due to this feature, the \xtranslate{} function is always \emph{total}, 
an important feature
when addressing domain experts, who are not language experts. 
An interesting philosophical question is the 
definition of the content model reported by \xmissing(): E.g.\ when the
original syntax requires something like ``\texttt{(a|b)?, x, d+}'', which 
is not matched by the input, then minimally only ``\texttt{x,d}'' is required 
(in the strict sense of the word) to make the input complete.
Nevertheless we decided to report the subexpression from the original structure
definition as a whole, to make the error more easily locatable by the user.

% -----------------------------------------------------------------------
\begin{table}[tp]
  \begin{align*}
  \begin{gathered}
\begin{array}{lcl}
    N &::=& \xnode(I, N^*)~~|~~\xchars(C^\star)~~|~~\xperm(T \times (T\pfun N^\star)) \\
      &|& \xmissing (T) ~~|~~ \xskipped(D)
\end{array}\\
\begin{array}{lcl}
T &::=& \xop'\ldots\xop' ~|~\xop"\ldots\xop"~|~\xop{0x}\ldots 
%%\\  &|&
~|~   T~\xop{..}~T ~|~ T~\xop U~T ~|~ T~\xop A~T ~|~ T~\xop-~T  \\
    &|&   T~\xop{\~{}} ~T ~|~ T~\xop{\~{}+}~|~ T~\xop{\~{}*} 
  ~|~ \xop>~~T~|~ \xop[~id~T~\xop]\\
    &|&  \xop{\#chars} ~|~  T~\xop,~T ~|~ T~\xop+ ~|~ T~\xop* ~|~ T~\xop? ~|~ T~\xop\&~T
             ~|~ T~\xop|~T  \\
   &|& T~\hat{}~\xop{(}T\xop/ id\xop) ~|~id
~|~\xop@~id 
%%\\ &|& 
~|~ \xop( T\xop) ~|~ \xop{\#empty} ~|~~\xop{\#none}\\
\end{array}\\
X_R 
\mbox{\it //~~many additional attributes: representation, docu, XSLT, etc.}\\
\xDefinition \hateq [  \mathit{tag} : I~~;~~ 
\mathit{kind} : \{\xtags, \xchars\} ~~;~~ 
\mathit{regExp} : T ~~;~~ 
\mathit{repr} :  X_R   ] \\[2ex]
  \xdefs : I \pfun  \xDefinition\\
  \texttree : D^\star \times I \fun N \\
  F = T \times N^\star \\
  \xtranslate : D^\star \times F^\star \fun   D^\star \times F^\star \\
\xfrac{\xtranslate (d, \langle(\xdefs(i)\mathit{.regExp}, \xes)\rangle) 
= (\langle\Xeof\rangle,\langle(\_,n)\rangle)}
{\texttree(d,i) = n}\\
  I_C = I \cup \{\Xchars\} \\
 \xskip, \xskip^C : D^\star \times F^\star \fun  D^\star \times F^\star \\
\xfirst : T \fun \xfinpower I_C \mbox{~~~\it// the ``first'' set of the regexp}\\
\xpotEps : T \fun \xboolean \mbox{~~~\it// whether regexp matches empty input}\\
\xascend_O : I_C \times F^\star \times F^\star  \fun F^\star\\
\xascend_C : I \times F^\star \times F^\star  \fun F^\star\\
\xdescend : T \times I_C \fun F^* \\
\xfrac{e\not= \xchars(\_)}
{\xskip^C (\xchars(x)\xhead \delta,  (t, \tau)\xhead\phi)
= \xskip^C (\delta,  (t,\tau\xtail \xskipped(\xchars(x)))\xhead\phi)\\
\xskip^C (e\xhead \delta, \phi) = e\xhead \delta, \phi \\
\xskip (d\xhead \delta,  (t, \tau)\xhead\phi)
= \xskip^C (\delta,  (t,\tau\xtail \xskipped(d))\xhead\phi)\\
}\\
\xfrac{\phi'=\xascend_C, (j, \phi,\xes)}
{\xtranslate ( \xCLOSE_j \xhead\delta, \phi )
= \begin{cases} \xtranslate (\delta, \phi') 
                     & \textrm{if}~~\phi'\neq\xes \\
                \xtranslate (\xskip( \xCLOSE_j\xhead\delta, \phi ))
                     & \textrm{otherwise} 
  \end{cases} 
}\\
\xfrac{\phi=(t, \_)\xhead \_ ~~~ \phi'=\xascend_O(j,\phi, \xes)}
{\xtranslate ( \xOPEN_j\xhead\delta, \phi )
 = \begin{cases} \xtranslate (\delta, \xdescend(t,j)\chop \phi) 
                      & \textrm{if}~~j \in \xfirst(t)\\
                 \xtranslate (\delta, \xdescend(t,j)\chop \phi' ) 
                      & \textrm{if}~~j \not\in \xfirst(t) \land \phi' \neq\xes\\
                 \xtranslate (\xskip( \xOPEN_j\xhead\delta, \phi ))
                      & \textrm{otherwise} 
   \end{cases} 
}\\
\xfrac{ f=(t, \nu)~~~\phi'=\xascend_O(\Xchars,f\xhead\phi,\xes)
~~~\Delta=\xchars(x) \xhead\delta}
{\multicolumn1l{\xtranslate (\Delta, f \xhead \phi )}
\\
 = \begin{cases} 
                 \xtranslate (\delta, (t, \nu\xtail\xchars(x))\xhead\phi) 
                      & \textrm{if}~~\Xchars \in \xfirst(t)\\
                 \xtranslate (\Delta, \phi')
                      & \textrm{if}~~\Xchars \not\in \xfirst(t)\land\phi'\neq\xes\\
                 \xtranslate (\xskip^C( \Delta,f \xhead \phi ))
                      & \textrm{otherwise} 
   \end{cases}
}\\
\xtranslate ( \xEMPTY_j \xhead \delta, \phi) 
= \xtranslate ( \xOPEN_j \xhead \xCLOSE_j \xhead \delta, \phi) \\
  \end{gathered}
  \end{align*}
\caption{Tag Parsing}
\label{tab_tagparse}
\end{table}
% -----------------------------------------------------------------------

% -----------------------------------------------------------------------
\begin{table}[tp]
  \begin{align*}
  \begin{gathered}
\xfrac{
t, t', t_n \in T ~~i \in I_C ~~ i'\in I \\
i\in\xfirst(t)~~~i\not\in\xfirst(t')~~~ \square\in\{\xop+, \xop*\}\\
p = t_1 \xop\& \ldots\xop\&  ~t~  \xop\& \ldots \xop\& ~t_n
}
{
\xdescend( (t \xop, t_2 \xop, \ldots\xop, t_n), i)
= \xdescend(t, i) \xtail ( (t_2 \xop, \ldots\xop, t_n), \xes)\\
\xdescend( (t' \xop, t_2 \xop, \ldots\xop, t_n), i)
= \xdescend( (t_2 \xop, \ldots\xop, t_n), i)\\
\xdescend( (t_1 \xop| \ldots\xop| t \xop| \ldots \xop| t_n), i)
 = \xdescend( t, i)\\
\xdescend( p, i) = \xdescend(t, i) \xtail (t, \xperm(p, \{\})) \\
\xdescend( t\square~, i) = \xdescend( t, i) \xtail (t\square,\xes) \\
\xdescend( t\xop?, i) = \xdescend( t, i)\\
\xdescend( i', i') = \langle(\xdefs(i')\mathit{.regExp} ,\xes)\rangle \\
\xdescend( \Xchars, \Xchars) = \xes\\
}\\
\\
\xfrac{
i, j\in I~~~i\not=j~~~t  \in T~~~ \square\in\{\xop+, \xop*\} \\
 S = (t_1 \xop, t_2 \xop, \ldots\xop, t_n)\in T
}
{
\xascend_C(i, (j,\nu)\xhead\phi, \tau) = 
\xascend_C(i, \phi, \langle\xnode(j, \nu\chop\tau)\rangle \\
\xascend_C(i, (i,\nu)\xhead(t, \mu)\xhead \phi, \tau) = 
(t, \mu\xtail\xnode(j, \nu\chop\tau))\xhead\phi\\
\xascend_C(i, (t\square,\nu)\xhead \phi, \tau) = \xascend_C(i, \phi, \nu\chop\tau)\\
\xascend_C(i, (S, \nu)\xhead \phi, \tau) = 
\begin{cases}
\xascend_C(i, \phi, \nu\chop\tau \xtail\xmissing(S))&\mathbf{if}~~\lnot \xpotEps(S) \\
\xascend_C(i, \phi, \nu\chop\tau )                  &\mathbf{otherwise}\\
\end{cases}
}%
\\
\\
\xfrac{
i \in I~~~t  = (t_1 \xop\& \ldots \xop\& t_n)\\
M = \{t_{k_1}\mapsto v_{k_1}, \ldots, t_{k_m}\mapsto v_{k_m} \} ~~~
M' = M \cup \{ t_x \mapsto \tau \} \\
\mathsf{matches}(t_y) = 
t_y \in \{t_1,\ldots,t_2\} \land  i \in \xfirst(t_y) 
\land t_y \not\in \dom M' \\
R = \{ t_k \in \{t_1\ldots t_n\} | t_k \not\in \dom M' \land \lnot\xpotEps (t_k) \} ~~~
R' = \_\xop\&\_ ~\limg R \rimg
}
{
\xascend_C(i, (t_x, \xperm(t, M)) \xhead\phi, \tau) \\
= \begin{cases}
\xascend_C(i, \phi, 
   \langle\xperm(t, M\cup\{t_x\mapsto \tau\xtail \xmissing(R')\})\rangle) 
&\mathbf{if}~~R \not= \{\}\\
\xascend_C(i, \phi,
   \langle\xperm(t, M')\rangle)               &\mathbf{otherwise}\\
\end{cases}
\\
\xascend_O(i, (t_x, \xperm(t, M)) \xhead\phi, \tau) \\
= \begin{cases}
(t_y, \xperm(t, M'))\xhead\phi 
&\mathbf{if}~~\exists t_y \bullet \mathsf{matches}(t_y) \\
\xascend_O(i, \phi, 
   \langle\xperm(t, M\cup\{t_x\mapsto \tau\xtail \xmissing(R')\})\rangle) 
&\mathbf{otherwise, if}~~R \not= \{\}\\
\xascend_O(i, \phi,
   \langle\xperm(t, M')\rangle)               &\mathbf{otherwise}\\
\end{cases}
\\
}\\
\\
\xfrac{
i, j\in I~~~t, t'  \in T~~~ \square\in\{\xop+, \xop*\} \\
i \in \xfirst(t)~~i \not\in \xfirst(t)\\
 S = (t_1 \xop, t_2 \xop, \ldots\xop, t_n) \in T
% ~~~
}
{
\xascend_O(i, (j,\nu)\xhead\phi, \tau) = 
\xascend_O(i, \phi, \langle\xnode(j, \nu\chop\tau)\rangle \\
\xascend_O(i, (t\square,\nu)\xhead \phi, \tau) = 
(t\square,\nu\chop\tau) \xhead \phi \\
\xascend_O(i, (t'\square,\nu)\xhead \phi, \tau) = \xascend_O(i, \phi, \nu\chop\tau)\\
\xascend_O(i, (S, \nu)\xhead \phi, \tau) \\
= \begin{cases}
((t_2~\xop,~\ldots~\xop,~t_n), \nu\chop\tau)\xhead \phi) 
&\mathbf{if}~~i \in \xfirst(t_1) \\
\xascend_O(i, ((t_2~\xop,\ldots\xop,~t_n), \nu)\xhead \phi,
\tau \xtail\xmissing(t_1))&\mathbf{otherwise, if}~~\lnot \xpotEps(t_1) \\
\xascend_O(i, ((t_2~\xop,\ldots\xop,~t_n), \nu)\xhead \phi,
\tau)&\mathbf{otherwise, if}~~n\geq 2\\
\xascend_O(i, \phi, \nu\chop \tau)&\mathbf{otherwise}\\
\end{cases}
}%
\\
  \end{gathered}
  \end{align*}
\caption{Tag Parsing, continued}
\label{tab_tagparse_II}
\end{table}
% -----------------------------------------------------------------------

% --------------------------------------------------------------------------------
\subsection{Parsing of XSLT Sources}

The algorithm is slightly enhanced to parse XSLT sources. In these,
elements (and attributes) of the XSLT language and those of the 
target language appear intermingled. Again, we want to write with least
noise, and both categories thus must be recognized automatically, as
far as possible.
These measures are taken:

\begin{itemize}\setlength{\itemsep}{0.0pt}\setlength{\parsep}{0.0pt}\setlength{\parskip}{0.0pt}
\item
Basically, the XSLT language and the target language must be provided as
text structure definitions. They will be parsed by switching between the
corresponding state machines, in the style of ``co-routines''.
\item
All reachable elements of the target language 
are collected, and all XSLT elements
which are allowed to contain target language elements. 
\item 
Whenever parsing the contents of an element from the latter set, 
an opening tag of the former set may appear, additionally to the normal parsing
as controlled by the XSLT grammar state machine.
\item
Vice versa, dedicated (``productive'') XSLT elements can contain 
anywhere in a target element. Whenever this happens, the parsing process
is switched to ``weak mode'', a variant, in which every target content
expression is allowed to additionally match the empty string, as if 
decorated with a ``\xop?''. 

A more promising approach will be the integration of \emph{Fragmented Validation (FV)}.
This is a technique of parsing the result fragments in an XSLT source
by a non-deterministic parser which follows all possible situations
in parallel. It has been presented in \cite{lt_xslt13}, but not yet
integrated into the \ddd{} tool.
\item
The tags of the target language have priority over the same tags from XSLT.
For these prefixed aliases are generated automatically, 
whenever necessary.
\end{itemize}

In practice it turned out that this format for writing down XSLT sources
allows a work-flow nearly as with a dedicated programming language front-end.

% -----------------------------------------------------------------------
\begin{table}[tp]
  \begin{align*}
  \begin{gathered}
S = \nat \times N^\star \\
\xparseM : T \times \xfinpower S ~~\fun~~\xfinpower S \\
\xparse : T \times S ~~\fun~~\xfinpower S \\
\xparseM (t, s) = \bigcup s' \in s \bullet \xparse (s') \\
\\
\xparse ( t, \{\} ) = \{\} \\
\xparse ( t \xop?, s) = \xparse(t, s) \cup \{s\} \\
\xparse ( t \xop{\~{}*}, s) = \xparseM(t\xop{\~{}*}, \xparse(t, s)) \cup \{s\} \\
\xparse ( t \xop{\~{}+}, s) = \xparseM(t\xop{\~{}*}, \xparse(t, s))  \\
\xparse ( t_1 \xop| t_2, s) = \xparse(t_1, s) \cup  \xparse(t_2, s) \\
\xparse ( \xop" \alpha \xop"~, (p, n)) = 
\begin{cases}
\{ (p + \mathsf{len} (\alpha), n) \} 
     & \mathbf{if} \mbox{\it ~~$\alpha$ is prefix of data at pos $p$} \\
\{\} & \mathbf{otherwise} 
\end{cases}\\
\xparse ( \xop' \alpha \xop'~, (p, n)) = 
\begin{cases}
\{ (p + 1, n) \} 
     & \mathbf{if} \mbox{\it character at pos $p$ is contained in $\alpha$} \\
\{\} & \mathbf{otherwise} 
\end{cases}\\
\\
\xparse(t_1 \xop| t_2, s) = \xparse(t_1, s) \cup  \xparse(t_2, s) \\
\\
\xparse(t_1 \xop, t_2, s) = \xparse(t_1 \xop{\~{} " "* \~{}} t_2, s)\\
\\
\xfrac{\xparse(t_1, s) = S
}
{\xparse(t_1~~\xop{\~{}}~~t_2, s) = \xparseM(t_2, S) \\
\xparse(\xop{>}~~t_1, s) = \mu~~\{ (p,r) \in S ~|~ (\forall (p',\_) \in S
\bullet p'\leq p)~\}\\
%%\xparse(\xop{>}~~t_1, s) = \{ (p,r) \in S ~|~ p \mbox{\it~is maximal in $S$}~\}\\
}
\\[6ex]
\xfrac{\xparse(t_1, (p, r)) = S \\
\mathit{pack} (p', r') = (p', r \xtail \xnode(i, \mathit{packRes}(p',r')) \\
\mathit{packRes}(p',r') = 
\begin{cases} \xchars (\mathit{data.substring}[p, p']) &\textbf{if}~~~ r' == \xes \\
              r'                                      &\textbf{otherwise} \\
\end{cases}
}
{\xparse(\xop[ i ~~t_1~\xop], (p, r)) = \mathit{pack} \limg S \rimg
} 
\\
  \end{gathered}
  \end{align*}
\caption{Character Based Parsing}
\label{tab_charparse}
\end{table}
% -----------------------------------------------------------------------

% --------------------------------------------------------------------------------
\subsection{Character Based Parsing}

Whenever the opening tag of an element has been consumed which is declared
as a character parser, the text input is redirected to the parsing process
as defined in Table~\ref{tab_charparse}.
Character based parsers have non-deterministic semantics: A set of hypotheses
is maintained in parallel, and at the end the longest matched prefix is
delivered.

The first operator special for the character level
is  ``\xop\~{}'', which is a sequential composition \emph{without}
intervening whitespace. This hold also for the repetition operators 
``\xop{\~{}*}'' and ``\xop{\~{}+}'' 
The operator ``\xop,''  is taken over from the tag level, for convenience,
and means sequential composition with arbitrary intervening whitespace.
The operator ``\xop\&'' for permutation is currently \emph{not} supported
on character level.
The operator ``\xop>'' defines a greedy sub-expression, where non-determinism
is overruled by longest prefix matching. The operators ``\xop{\~{}*}'' and 
 ``\xop{\~{}*}'' are greedy anyhow, when they are applied to plain 
character sets.

The generated and output XML expression will be one single element, with the
parser's identifier as its tag, if no structure is defined.
This is done by nesting the constructs ``$\xop[~i~T~\xop]$'', which generate an
element with $i$ as its tag and the parsed contents. This again is simply
the parsed character data, if no such constructs are contained recursively,
or the sequence of the resulting XML elements, otherwise.

Immediately after the character parser cannot be continued, control returns
to the tag parsing level. Now an explicit closing tag for the parser may follow,
but is never required.

% --------------------------------------------------------------------------------
\section{Text Format Definitions}
\label{txtDdf}

The genuine \ddd{} text structure definition language ``ddf'' supports (a)
the definition of the element content models, as tag or character parsers,
plus (b) various additional parameters.

% --------------------------------------------------------------------------------
%\subsection{Definition of Parsers}

The content models follow basically the same design principles as known
from W3C DTD \cite{xml11} or relaxNG \cite{relaxng}. 
Differences are the ``\xop\&'' operator, which does not stand for
interleaving (as in relaxNG), but only for permutation.
New is the ``\xop@'' operator, which inserts the content model of the 
referred definition into any expression, allowing a definition to act as
element specification and as mere internal constant (with an expression as its
value).

% --------------------------------------------------------------------------------
%\subsection{Additional Modifiers}

Each definition may carry attributes related to very different layers:
The XML tag can be set, overriding the identifier of the definition, which is the
default; an XML namespace URI can be defined; different formats for editing
and parsing can be specified; definitions can open a local scope 
for tags; etc.

More importantly: User documentation in different languages can be attached
to every definition, 
and XSLT rules for transformation into different back-ends.
Both features employ \ddd{} recursively to document or process itself: 
The former employs some standard format for technical documentation,
readable by humans; the latter employs the XSLT source format, as described
above, instantiated with the text structure definition of the target format.

All definitions are organized in a hierarchy of modules; each top module
must be locatable by some rules matching the module name to a file location, or sim.

Modules can be \emph{imported} into other modules. Thereby an ``import key''
is defined,
which used as a prefix makes the definitions in the imported module accessible in 
all expressions in the importing module. This includes ``automatic re-export'':
the imports in the imported module are accessible by concatenating these
import keys, etc.

Additionally, \emph{substitutions} can be defined which apply to all expressions
in the imported module (or only the expression of one particular definition).
Each such substitution replaces a particular reference, i.e.\ a sequence
of identifiers meant as a reference to a definition, by a given expression,
evaluated in the context of the importing module.
Furthermore, a particular module import in the imported module can be replaced
by a different one, defined in the importing module, as a whole.
This selection of mechanisms for parametrization has turned out to 
be very powerful and adequate when maintaining and developing mid-scale
text structure architectures, like the examples listed at the beginning.

\vspace*{-0.5cm}
\bibliographystyle{plain}
\bibliography{../lib/own2011,../lib/own2013,../ddd2011/d2d.bib}

\end{document}